\documentclass[aps,prl,showpacs,twocolumn,amsmath,amssymb,superscriptaddress,footinbib]{revtex4}

\usepackage[english]{babel}
\usepackage{latexsym}
\usepackage{graphics}
\usepackage{subfigure}
\usepackage{epsfig}
\usepackage{amsfonts}
\usepackage{amssymb}
\usepackage{amsmath}
\usepackage{bbm}
\usepackage{color}

\newcommand{\via}{{\it via}~}

\newcommand{\beq}{\begin{equation}}
\newcommand{\eeq}{\end{equation}}

\newcommand{\bea}{\begin{eqnarray}}
\newcommand{\eea}{\end{eqnarray}}

\newcommand{\ket}[1]{\ensuremath{|#1\rangle}}

\newcommand{\braket}[2]{\ensuremath{\langle #1|#2\rangle}}
\newcommand{\ew}[1]{\ensuremath{\langle #1\rangle}}

\newcommand{\eins}{\ensuremath{\mathbbm 1}}

\newcommand{\jhat}{\ensuremath{\hat{\jmath}}}

\newcommand{\vecb}[1]{\mbox{\boldmath$#1$}}

\begin{document}

\title{Quantum polarization spectroscopy of ultracold spinor gases}

\author{K. Eckert}
\affiliation{Departament F\' isica, Grup de F\' isica Te\'orica,
Universitat Aut\'onoma de Barcelona, E-08193 Bellaterra, Spain.}
\author{\L. Zawitkowski}
\affiliation{Centrum Fizyki Teoretycznej, Polska Akademia Nauk, Warszawa 02668, Poland.}
\author{A. Sanpera}
\affiliation{ICREA: Instituci\'o Catalana de Recerca i Estudis Avan\c cats.}
\affiliation{Departament F\' isica, Grup de F\' isica Te\'orica,
Universitat Aut\'onoma de Barcelona, E-08193 Bellaterra, Spain.}
\author{M.~Lewenstein}
\affiliation{ICREA: Instituci\'o Catalana de Recerca i Estudis Avan\c cats.}
\affiliation{ICFO--Institut de Ci\`encies Fot\`oniques,
E-08860 Castelldefels, Barcelona, Spain.}
\author{E.S.~Polzik}
\affiliation{ICFO--Institut de Ci\`encies Fot\`oniques,
E-08860 Castelldefels, Barcelona, Spain.}
\affiliation{Niels Bohr Institute, Danish Quantum Optics Center – QUANTOP, Copenhagen
University, Blegdamsvej 17, Copenhagen 2100, Denmark.}

\date{\today}

\begin{abstract}
We propose a method for the detection of ground state quantum
phases of spinor gases through a series of two quantum
nondemolition measurements performed by sending off-resonant,
polarized light pulses through the gas. Signatures of various
mean-field as well as strongly-correlated phases of $F=1$ and
$F=2$ spinor gases obtained by detecting quantum fluctuations
and mean values of polarization of transmitted light are identified.
\end{abstract}


\maketitle


It has been demonstrated several years ago
that fundamental quantum spin noise of a
collection of cold atoms can be measured {\it via} quantum noise limited
polarization spectroscopy \cite{sorensen:1998}. Since then, a quantum interface of light with optically
thick atomic spin ensembles has become  a promising and
powerful method for a transfer of quantum information between atomic
internal degrees of freedom and light.  The
basic concept underlying such atom-light interfaces is provided by
the off-resonant coupling of the collective atomic spin, i.e., of
several magnetic sublevels, to the polarization of light. In particular,
such an off-resonant interaction,
followed by a quantum measurement on light, has been shown to be
a powerful quantum nondemolition (QND) tool to
generate spin squeezed and entangled atomic states
\cite{kuzmich:1998,kuzmich:2000,mabuchi:2004,julsgaard:2001,julsgaard:2004},
to teleport quantum states
between ensembles \cite{duan:2000}, or to propose
\cite{kozhekin:2000,lukin:2002} and realize
\cite{julsgaard:2004} high fidelity quantum memories for light.

In the present paper we propose to apply quantum
polarization spectroscopy techniques for the detection of various
quantum phases of degenerate atomic gases, i.e., with atoms having spin degrees of freedom.
Such ultracold spinor gases have recently brought a new perspective to the study of magnetic systems. Seminal experiments
of the MIT group with an optically trapped spin $F=1$ Sodium condensate \cite{stamperkurn:1998}, and
theory papers of Ho \cite{ho:1998}, and Ohmi and Machida \cite{ohmi:1998} have triggered the
use of cold atoms to study magnetic ordering and domains.  Spin interaction effects
are very much enhanced in the strongly correlated
regime, which nowadays is reachable experimentally \cite{mandel:2003} by loading an ultracold
spinor gas into an optical lattice so that the kinetic energy (tunneling)
becomes small in comparison with atom-atom interactions. Then the atoms can be well
described by a generalized spinor Bose-Hubbard Hamiltonian (BHH) \cite{imambekov:2003, yip:2003,zawit:2006}.
In the limit of small occupation number it reproduces accurately
(within the experimentally achievable regime) some of the most paradigmatic spin
chain models.
Experimental observation of the rich variety of magnetic ordering
present in these systems remains however elusive due to similiar values of the scattering lengths on the different spin collision channels.

A way to determine properties of a quantum spinor gas could be to use a strong QND measurement.
As shown here, 
a series of QND measurements using
polarization spectroscopy on light transmitted through a
condensate yields mean values and variances of the atomic total spin
operators, thus allowing unambiguous distinction
of various atomic quantum phases.


{\it Formalism --} We consider a sample of neutral atoms in a $2F+1$--
dimensional ground state manifold $\ket{F,m}$, interacting
off-resonantly with linearly polarized light propagating along the
$z$-direction (cf. \cite{julsgaard:2003:thesis}).
After adiabatically eliminating excited atomic states, the interaction 
can be described {\it via} an effective Hamiltonian
\bea\label{eqn:heff}
\hat{H}_{\textrm{int}}^{\textrm{eff}}=
-\!\int_0^{L}\!\!\!\!\!dz\rho A
(
    a_0\hat{\phi}+
    a_1\hat{s}_z\hat{\jmath}_z+
a_2[
        \hat{\phi}\hat{\jmath}_z^{2}-
        \hat{s}_-\hat{\jmath}_+^{2}-
        \hat{s}_+\hat{\jmath}_-^{2}
    ]
), \eea
where $L$ is the length of the atomic sample.
The conditions under which decoherence due to absorption of light
can be neglected, such that this Hamiltonian is valid, will be
discussed below.
In Eq.~(\ref{eqn:heff}),
$a_i\propto\hbar\gamma\lambda^2 c/(16\pi A\Delta)$, where $A$ is the cross-section of the
atomic sample overlapping with the probe light, $\Delta$ is the detuning, $\rho$ is the (in general $z$-dependent)
atomic density, $\lambda$ is the wavelength, and $\gamma$ is the
excited state line width.
$\hat{s}_{\alpha}\equiv\hat{s}_{\alpha}(z,t)$
($\hat{s}_{\pm}=\hat{s}_x\pm i\hat{s}_y$)
are the components of the Stokes vector characterizing the
polarization of the light pulse, $\hat{\phi}(z,t)$ is the photonic
density, and $\hat{\vecb{\jmath}}\equiv\hat{\vecb{\jmath}}(z,t)$
are atomic spin operators.  The term proportional to
$a_0$ corresponds to the AC Stark shift, while
$a_2\rightarrow0$ for values of $\Delta$ large
compared to the excited state hyperfine structure
\cite{julsgaard:2003:thesis}. Here we assume $a_2=0$ and restrict to the
linear coupling between the Stokes operator and the atomic spin, which
represents a QND Hamiltonian.
Using Heisenberg equations of motion, we find that
$\hat{\jmath}_z$ is conserved. For a pulsed probe with
a duration of $\mu$s we can thus assume
$\hat{\jmath}_z(z,t)\equiv\hat{\jmath}_z(z)$, as spin diffusion
happens on a much larger timescale of ms. 
For a probe strongly polarized along the $x$-direction, the macroscopic Stokes
operator $\hat{S}_x$ ($\hat{S}_{\alpha}=\int dt\hat{s}_{\alpha}$)
can be replaced by a c-number
$\hat{S}_x\approx\ew{\hat{S}_x}=N_{P}/2$, being $N_{P}$ the number of photons.
Neglecting retardation effects,
the propagation equation for $\hat{S}_y$ reads $\partial_z \hat{S}_y(z)=-a_1A\rho\ew{\hat{S}_x}\jhat_z(z)$.
It is convenient to introduce the collective spin in $z$-direction $\hat{J}_z=\int_0^{L}dz\,A\rho
\jhat_z(z)$ and to define quadrature operators
\via $\hat{X}_S=\sqrt{2/N_{\rm P}}\hat{S}_y$, $\hat{Y}_S=\sqrt{2/N_{\rm P}}\hat{S}_z$,
such that $[\hat{X}_S,\hat{Y}_S]\approx i$.
Integrating the propagation equation gives
\beq
 \hat{X}_S^{\rm out}=\hat{X}_S^{\rm in}-\frac{\kappa}{\sqrt{FN_A}}\,\hat{J}_z,\label{eqn:evol_s}
\eeq 
with $\kappa^2=a_1^2N_{\rm P}N_A\frac{F}2$ (number of
atoms $N_A=\int dz\,\rho A$). Eq.~(\ref{eqn:evol_s}) is valid provided that
$2\kappa^2FN_A/N_P\ll1$.
Fluctuations of the $\hat{X}_S$
quadrature are calculated as 
\bea\label{eqn:quadgen} (\Delta
\hat{X}_S^{\rm out})^2&=&\frac12+\frac{\kappa^2}{FN_A}
\int_0^L dz\int_0^L dz'\rho^2A^2\nonumber\\
&&\mspace{-75mu}\times\left[\ew{\jhat_z(z)\jhat_z(z')}-\ew{\jhat_z(z)}\ew{\jhat_z(z')}\right],
\eea 
where $1/2$ stems from the quantum fluctuations
of the coherent input state of light (preparing a squeezed input state
would allow to reduce this contribution).

Equations (\ref{eqn:heff}--\ref{eqn:quadgen}) allow to
estimate the feasibility of the proposed methods for measurement of
the fluctuations of $\hat{J}_z$ of a spinor condensate. A
strong QND measurement repeated on samples prepared in the same
state yields a complete knowledge about the operator $\hat J_z$.
The strength of the interaction is
determined by the constant $\kappa$. In order to obtain useful
information about the spin fluctuations, the second term in
Eq.~(\ref{eqn:quadgen}) must be large compared to $1/2$, the
quantum noise of a coherent probe. For the sake of this estimate
we take a ferromagnetic spinor condensate (also
referred to as a coherent spin state outside the context of
quantum gases), for which the second term is equal to $\kappa^2/2$
\cite{kuzmich:2000,mabuchi:2004}. The interaction
constant can be expressed \cite{interface:2004} as
$\kappa^2=\alpha\eta$, where $\eta$ is the probability of
spontaneous excitation caused by the off-resonant probe and
$\alpha$ is the resonant optical depth of the sample. $\eta$
describes the decoherence which must be kept small by choosing the
detuning and/or the strength of the probe pulse, in order to
minimize the distortion of $\hat{J}_z$ and to validate the use of
Hamiltonian Eq.~(\ref{eqn:heff}). 
The optimal value for a QND measurement
of $\hat{J}_z$ is $\kappa^2=\sqrt{\alpha/2}$ \cite{interface:2004}. Hence for an
optically thick spinor condensate with $\alpha\gg1$ the proposed
method works well. For degenerate gases, typical values of $\alpha=300$ or even more
have been demonstrated \cite{varenna:1999}.

The most general Hamiltonian describing a spinor atomic sample is
given by: \beq
\hat{H}_{at}\!=\!\int\!\!d\vecb{r}\!\!\!\sum_{m=-F}^F\!\!\hat{\Psi}_m^{\dagger}(\vecb{r})\!\!
\left(\!-\frac{\hbar^2\Delta^2}{2M}+U_{\rm trap}(\vecb{r})\!\right)\!\hat{\Psi}_m(\vecb{r})+\hat{V}_{\rm
int},
\label{hamiltonian}
\eeq
where the field operator $\hat{\Psi}_m^{\dagger}(\vecb{r})$ creates a particle with
spin projection $m$ at position $\vecb{r}$ and $\hat{V}_{\rm int}$ describes the atom-atom interactions.
For two identical spin-$F$ bosons interacting via $s$-wave collisions, $\hat{V}_{\rm int}=\sum_{M=0}^Fg_{2M}\hat{P}_{2M}$,
with $\hat{P}_{2M}$ being the projector onto the subspace with total spin $F^{\rm tot}=2M$,
and $g_{2M}$ the interaction strength for the given spin channel.
For a spatially uniform trapping potential,
condensation occurs in the zero momentum state,
and variational ground states are obtained from a trial wavefunction \cite{ueda:2002}
\beq
\ket{\xi}=\frac1{\sqrt{N_A!}}\left[\sum_{m=-F}^{F}\xi_m \hat{a}_{\vecb{k}=0,m}^{\dagger}\right]^{N_A}\ket{\rm vac},
\eeq
with the complex components of the vector $\vecb{\xi}$ as parameters. $\hat{a}_{\vecb{k}=0,m}^{\dagger}$
creates a particle with spin projection $m$ in the $\vecb{k}=0$ state.
Ground states have been discussed
in detail for $F=1$ \cite{ho:1998}, $F=2$ \cite{ueda:2002}, and $F=3$
\cite{diener:2005}. Given the spinor $\vecb{\xi}$, means and variances
are calculated as
\begin{eqnarray}\label{eqn:meanvarMF}
\ew{\hat{X}_S^{\rm out}}\!\!&=&\!\!-\frac{\kappa\sqrt{N_A}}{\sqrt{F}}\sum_{m=-F}^Fm|\xi_m|^2,\\
(\Delta \hat{X}_S^{\rm out})^2\!\!\!&=&\!\!\frac12\!+\!\frac{\kappa^2}{F}
\!\!\left[\sum_{m=-F}^{F}\!\!\!m^2|\xi_{m}|^2\!-\!\!\left(\sum_{m=-F}^{F}\!\!\!m|\xi_{m}|^2\!\right)^{\!\!2}\right].\nonumber\label{eqn:varMF}
\end{eqnarray}

\begin{figure*}[t]
\includegraphics[width=0.9\linewidth]{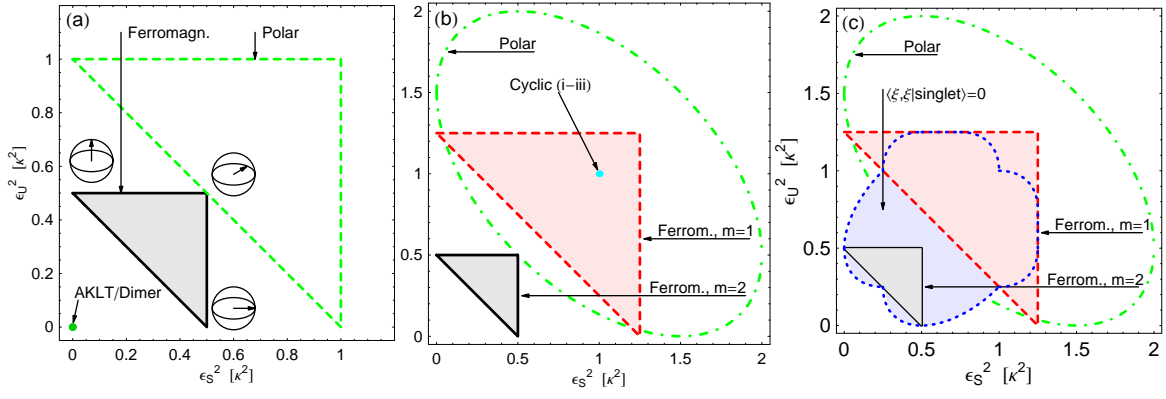}
\caption{
(color online) Possible combinations of additional fluctuations $\epsilon_S^2$ and $\epsilon_U^2$ imprinted on the light
for the ground state phases of the $F=1$ spinor gas in a uniform trap and in an optical lattice (a)
and for $F=2$ atoms in a uniform trap (b)
and in an optical lattice (c). Filled areas denote cases where the mean of $\ew{X_{S}^{\rm out}}$ and/or
$\ew{\hat{X}_{U}^{\rm out}}$ is (generically) non-zero. The spheres in (a) illustrate the directions
of the spinor for the extremal points of the ferromagnetic phase. 
}
\label{fig:varall}
\end{figure*}

\begin{table*}[bt]
\begin{tabular}{|l|}
\hline
(ia) Ferromagnetic $m=2$\ \ \ --\ \ \ $\vecb{\xi}=(e^{-i2\alpha}s_{\beta}^4,2e^{-i\alpha}s_{\beta}^3c_{\beta},\sqrt{\frac34}S_{\beta}^2,2e^{i\alpha}s_{\beta}c_{\beta}^2,e^{i\alpha}c_{\beta}^4)$\\
$\ew{\hat{X}_S^{\rm out}}=-\sqrt{2N}C_{\beta},\,\,\,\ew{\hat{X}_U^{\rm out}}=-\sqrt{2N}S_{\alpha}S_{\beta}$,\ \ \
$\epsilon_S^2=\frac12\kappa^2S^2_{\beta},\,\,\,\epsilon_U^2=
\frac12\kappa^2\left(1-S^2_{\alpha}S^2_{\beta}\right)$\\\hline\hline
(ib) Ferromagnetic $m=1$\ \ \ --\ \ \
$\vecb{\xi}=(2e^{-i\alpha}s_{\beta}^3c_{\beta},-e^{-i\alpha}s_{\beta}^2(1+2C_{\beta}),-\sqrt{\frac23}S_{\beta}C_{\beta},-e^{-\alpha}c_{\beta}^2(-1+2C_{\beta}),2e^{\-2i\alpha}s_{\beta}c_{\beta}^3)$\\
$\ew{\hat{X}_S^{\rm out}}=-\frac{\sqrt{N}}2C,\,\,\,\ew{\hat{X}_U^{\rm out}}=-\frac{\sqrt{N}}2S_{\alpha}S_{\beta}$,\ \ \
$\epsilon_S^2=\frac54\kappa^2S^2_{\beta},\,\,\,
\epsilon_U^2=\frac54\kappa^2\left(1-S^2_{\alpha}S^2_{\beta}\right)$\\\hline\hline
(ii) Polar\ \ \ --\ \ \ $\vecb{\xi}=(e^{i\phi}S_{\theta}S_{\psi},e^{i\chi}S_{\theta}C_{\psi},\sqrt2C_{\theta},-e^{-i\chi}S_{\theta}C_{\psi},e^{-i\phi}S_{\theta}S_{\psi})$\\
$\ew{\hat{X}_S^{\rm out}}=0,\,\,\,\ew{\hat{X}_U^{\rm out}}=0$,\ \ \
$\epsilon_S^2=\frac12\kappa^2S^2_{\theta}(1+3S^2_{\psi}),\,\,\,
\epsilon_U^2=\frac12\kappa^2\left(3-2S^2_{\psi}+3C^2_{\chi}C^2_{\psi}C^2_{\theta}-
\sqrt{3}C_{\phi}S_{2\theta}S_{\psi}\right)$\\\hline\hline
(iii) Cyclic\ \ \ --\ \ \ as \cite{ueda:2002}, Eq.~(60--62), for zero magnetic field:
\ \ $\ew{\hat{X}_S^{\rm out}}=0,\,\,\,\ew{\hat{X}_U^{\rm out}}=0$,\ \ \
$\epsilon_S^2=\kappa^2,\,\,\,
\epsilon_U^2=\kappa^2$\\\hline
\end{tabular}
    \caption{Spinors $\vecb{\xi}$ for the mean field ground state phases in the spin-$2$ case \cite{ueda:2002} and corresponding results for means $\ew{\hat{X}_{S/U}^{\rm out}}$ and additional fluctuations $\epsilon_{S/U}$.
Abbreviations $s_{x}=\sin\frac{x}2,\, c_{x}=\cos\frac{x}2,\,S_{x}=\sin x,\,C_{x}=\cos x$ are used.}
    \label{tab:Spin2MeanField}
\end{table*}
In the strongly correlated regime, Eq.~(\ref{hamiltonian}) reduces to a spinor
BHH.
For an integer filling factor  and
considering tunneling perturbatively up to second order,
an effective Hamiltonian with nearest neighbor spin-spin interaction arises.
For such a situation, ground state configurations have been analyzed 
in \cite{imambekov:2003} for $F=1$ and in \cite{zawit:2006,barnett:2006} for $F=2$. 
Eq.~(\ref{eqn:quadgen}) reduces then to a sum over correlations between all pairs of atoms:
\beq
(\Delta \hat{X}_S^{\rm out})^2=\frac12+\frac{\kappa^2}{FN_A}
\sum_{k,l=1}^{N_A}\left[\ew{\jhat^k_z\jhat^l_z}-\ew{\jhat_z^k}\ew{\jhat_z^l}\right].
\eeq
Notice that in $\hat H_{\rm int}^{\rm eff}$, Eq.~(\ref{eqn:heff}), light couples to individual atoms.
Thus for more than one atom per site,
$\vecb{\hat{\jmath}}^k$ denotes the
spin operators of the $k$th atom, {\it not} the total spin at site $k$.

The procedure described so far allows to obtain mean and variance of $\hat{J}_z$.
Reading an orthogonal component of the atomic spin vector allows for a
better discrimination of distinct phases.
Since the interaction with the first light pulse modifies the atomic spin state,
here we assume to have at our disposal a second,
identically prepared atomic sample. A lightpulse
incident in the $y$-direction, again strongly $x$-polarized, corresponds to
exchange $\jhat_z\rightarrow\jhat_y$, $\jhat_y\rightarrow-\jhat_z$
in Eqs.~(\ref{eqn:heff}--\ref{eqn:quadgen}). Thus it allows to read out the $y$-component
of the atomic spin vector, with the corresponding quadrature denoted as $\hat{X}_U$.



{\it Detecting spin-1 quantum phases --}
In the mean field case, a gas of $F=1$ atoms has two possible ground states
\cite{ho:1998}: a ferromagnetic one with all spins having maximal spin projection
in some direction $(\cos\alpha\sin\beta,\sin\alpha\sin\beta,\cos\beta)$, and thus $\ew{\vecb{\hat{\jmath}}}\neq 0$, 
and a polar one with $\ew{\vecb{\hat{\jmath}}}=0$.
Inserting the ferromagnetic spinor $\vecb{\xi}_{\rm f}=(e^{-i\alpha}\cos^2\frac{\beta}2,\sqrt2\cos\frac{\beta}2\;\sin\frac{\beta}2,
e^{i\alpha}\sin^2\frac{\beta}2)$ into Eqs.~(\ref{eqn:meanvarMF}) leads to
$\ew{\hat{X}_S^{\rm out}} = -\kappa\sqrt{N_A}\cos\beta$, 
$\ew{\hat{X}_U^{\rm out}} = \kappa\sqrt{N_A}\sin\alpha\sin\beta$,
$(\Delta \hat{X}_S^{\rm out})^2 = (1+\kappa^2\sin^2\beta)/2$, 
$(\Delta \hat{X}_U^{\rm out})^2 = (1+\kappa^2[1-\sin^2\beta\sin^2\alpha])/2$.
These equations are valid provided the mean
polarization of the probe light remains to be $x$-polarized which
is true under the feasible assumption $N_A\ll N_P$. For
the polar phase, the spinor can be parameterized as
$\vecb{\xi}_{\rm p}=(-e^{-i\alpha}\sin\beta,
\sqrt2\cos\beta,e^{i\alpha}\sin\beta)/\sqrt2$, and we have
$\ew{\hat{X}_S^{\rm out}}=0=\ew{\hat{X}_U^{\rm out}}$ and
$(\Delta \hat{X}_S^{\rm out})^2 = 
(1+2\kappa^2\sin^2\beta)/2$,
$(\Delta \hat{X}_U^{\rm out})^2 =
(1+2\kappa^2[1-\sin^2\alpha\sin^2\beta])/2$.
Characterizing the atomic phases by the additional noise $(\Delta \hat{X}_{S/U}^{\rm out})^2-\frac12\equiv\epsilon_{S/U}^2$
imprinted on the $\hat{X}$ quadratures, we obtain for the ferromagnetic phase
$\epsilon_U^2=\frac12\kappa^2-\epsilon_S^2\sin^2\alpha$,
with $0\leq\epsilon_S^2\leq\frac12\kappa^2$, and for the polar phase
$\epsilon_U^2=\kappa^2-\epsilon_S^2\sin^2\alpha$,
with $0\leq\epsilon_S^2\leq\kappa^2$. Possible values of the additional noise
lie in non-overlapping triangles in the $(\epsilon_S^2,\epsilon_U^2)$-plane,
see Fig.~\ref{fig:varall} (a).
Thus both phases can be distinguished through the noise imprinted on the light.
%
In this particular case, this is also possible by comparing the mean
values $\ew{\hat X_{U/S}}$ (for $(\alpha,\beta)\neq(0,\pi/2)$).


For the $F=1$ lattice gas with a single particle per site, the effective
Hamiltonian is
$\hat{H}_{\rm lat}=\sum_{\ew{kl}}(\cos\delta\,\vecb{\hat{\jmath}}^k\vecb{\hat{\jmath}}^l+\sin\delta\,(\vecb{\hat{\jmath}}^k\vecb{\hat{\jmath}}^l)^2)$, 
where the sum runs over nearest neighbors.
%
We consider only the ground states
of $^{23}$Na discussed by Imambekov {\it et al.} \cite{imambekov:2003} (for numerical investigations
see \cite{rizzi:2005}): (i) for a fully polarized state the properties of the
out-going light are as in the ferromagnetic case discussed before; (ii) a state mixing total
spin $F^{\rm tot}=0$ and $F^{\rm tot}=2$ on each bond, constructed as $\prod_m\ket{\vecb{\xi}_{\rm p}}_m$,
gives results as for the polar mean field state;
(iii) singlets  (dimers) can be put on every second bond (in 1D), breaking
translational symmetry. The reduced on-site density matrix
is $\rho_{\rm k}=\eins/3$, and thus $\ew{X_S^{\rm out}}=0$. As
$\ew{\jhat_k^z\jhat_l^z}=\frac23$ for $k=l$, $-\frac23$ for nearest-neighbors in a singlet state, and $0$ otherwise,
for an even number of sites that the fluctuations are unchanged:
$(\Delta \hat{X}_S^{\rm out})^2=1/2$. For an odd number of sites in 1D, or
randomly oriented dimers in 2D, $n$ atoms will be unpaired and thus
$
(\Delta \hat{X}_S^{\rm out})^2=1/2+2{n}\kappa^2/(3N_A)
$.
Due to the rotational symmetry of the singlet, the same result is
obtained for the $\hat{X}_U$ quadrature. (iv) In 1D and for $\delta=-\arctan(1/3)$,
the ground state is a valence bond solid (VBS) state \cite{aklt:1987}.
In this case, two-site correlations decay as
$\ew{\jhat_z^k\jhat_z^l}=\frac43(-\frac13)^{|k-l|}$ ($k\neq l$) \cite{aklt:1987}. Since
the VBS is non-magnetized, the means of the quadrature components remain
unchanged. As the sum of $\ew{\jhat^k_z\jhat^l_z}$ over all pairs of atoms gives $2/3$
independent of the number of sites, there is no detectable change in the fluctuations:
$
(\Delta \hat{X}_S^{\rm out})^2=\frac12+2{\kappa^2}/3{N_A}
$.
Thus distinguishing between a dimer and a VBS state is difficult with this method,
but in principle possible for small $N_A$.

For two atoms per lattice site, in the limit of vanishing
tunneling the ground state consists of non-interacting singlets on each site.
As tunneling is increased, on-site states with total spin $2$ become
important. Using a variational ansatz $\prod_k \ket{\psi}_k$
with $\ket{\psi}_k=\cos\Theta\ket{F^{\rm tot}=0,F^{\rm tot}_z=0}+
\sin\Theta\ket{F^{\rm tot}=2,F^{\rm tot}_z=0}$, a
sharp jump of $\sin\Theta$ from $0$ to a non-zero value
is found as tunneling is increased \cite{imambekov:2003}. Evaluating the quadrature operators
of the outgoing light, we find
$\ew{\hat{X}_S^{\rm out}}=0=\ew{\hat{X}_U^{\rm out}}$, but modified noise properties:
$
(\Delta \hat{X}_S^{\rm out})^2=(1+3\kappa^2\sin^2\Theta\sin^2\beta)/2$,
$(\Delta \hat{X}_U^{\rm out})^2=(1\!+\!3\kappa^2\sin^2\!\Theta[1\!\!-\!\!\sin^2\alpha\sin^2\beta])/2$,
where $\alpha$, $\beta$ parametrize the direction of the component with $F^{\rm tot}=2$.
The sharp change in the nature of the ground state manifests clearly in the noise properties
of the outgoing light. Let us emphasize that, as light couples to single atoms, here the fluctuations
are different from those arising from fully polarized $F=2$ atoms, as will be discussed now.




{\it Detecting spin-$2$ quantum phases --} For $F=2$, there are three different ground state phases
in the mean field case \cite{ueda:2002}: (i) ferromagnetic,
$\ew{\vecb{\hat{\jmath}}}\neq 0$, characterized by spin projection $m=2$ (a) or $m=1$ (b) in some direction, (ii) polar,
characterized by $\sigma=\sum_m(-1)^m\xi_m\xi_{-m}\neq0$ and,
$\ew{\vecb{\hat{\jmath}}}=0$, and (iii) cyclic, having $\sigma=0$ and
$\ew{\vecb{\hat{\jmath}}}=0$. Our results are summarized in table \ref{tab:Spin2MeanField} and Fig. \ref{fig:varall}(b). 
Ferromagnetic and polar phases can be perfectly distinguished while the cyclic phase lies within the polar one.


Finally, we consider the $F=2$ lattice gas with a single particle per site,
following the discussion in \cite{zawit:2006}.
The effective Hamiltonian contains up to the fourth power of the Heisenberg interaction:
$\hat{H}_{\rm lat}=\sum_{\ew{kl}}\sum_{\alpha=1}^4\lambda_\alpha(\vecb{\hat{\jmath}}^k\vecb{\hat{\jmath}}^l)^\alpha$.
We will discuss here only some of the possible ground state phases arising for various combinations of $\lambda_{\alpha}$:
(i) for products of either ferromagnetic or polar on-site states results are as in the mean field case;
(ii) an anti-ferromagnetic state $\ket{\vecb{\xi}_m,\vecb{\xi}_{-m},\vecb{\xi}_m,\ldots}$,
with $\ket{\vecb{\xi}_m}$ ($\ket{\vecb{\xi}_{-m}}$) being the on-site state with spin projection $m$ ($-m$)
in some direction. As such {\it Ne{\'e}l states} are non-magnetized,
the means of the outgoing $\hat{X}$ quadratures
remain unchanged, but fluctuations are as for ferromagnetic states.
Ne{\'e}l states can thus be distinguished from ferromagnetic ones due to different means,
but cannot always be distinguished from polar states. (iii) certain combinations of $\lambda_{\alpha}$ favor product
states $\ket{\vecb{\xi},\vecb{\xi},\ldots}$ fulfilling
$\braket{\rm singlet}{\vecb{\xi},\vecb{\xi}}=0$, corresponding to $\sigma=0$.
The cyclic states discussed in the mean field case are particular instances of this phase, which however
is larger because the spin projection is not restricted to zero.
Possible combinations of fluctuations for the different phases are summarized in Fig.~\ref{fig:varall} (c).



{\it Conclusions} -- We have presented a method to detect different magnetic orders
arising in ultracold spinor gases.
it is based on an atom-light interface, realized through the off-resonant interaction
of a strongly polarized pulse of light and an ultracold ensemble.
Fluctuations imprinted on the quadratures
of the outgoing light contain information on the total atomic spin components. By comparing the
fluctuations arising when the light impinges on the atomic sample from different directions,
most ground state phases can be distinguished. We plan to apply our analysis to  the
complete classification of the mean field states for arbitrary $F$ presented recently
by Barnett {\it et al.} \cite{barnett:2006}.

We thank  K. Bongs, E. Demler, A. Polls, and K. Sengstock for discussions.
We acknowledge support from DPG (SFB 407, SPP 1116),
EU  Programmes ``SCALA'', "COVAQIAL", "QAP", ESF Network QUDEDIS, and Spanish MEC 
contracts FIS 2005-04627, -01369, and EX 2005-0830.




\end{document}